# Can the Existence of Dark Energy Be Directly Detected?

Martin L. Perl
*Stanford Linear Accelerator Center and Kavli Institute for Particle Astrophysics and Cosmology
Stanford University, 2575 Sand Hill Rd. Menlo Park, CA 94025, USA*

The majority of astronomers and physicists accept the reality of dark energy and also believe that it can only be studied indirectly through observation of the motions of stars and galaxies. In this paper I open the experimental question of whether it is possible to directly detect dark energy through the presence of dark energy density. Two thirds of this paper outlines the major aspects of dark energy density as now comprehended by the astronomical and physics community. The final third summarizes various proposals for direct detection of dark energy density or its possible effects. At this time I do not have a fruitful answer to the question: Can the Existence of Dark Energy Be Directly Detected?

## 1. DARK ENERGY DETECTION: AN EXPERIMENTAL PROBLEM

### 1.1. Magnitude of Dark Energy Density

Many physicists and astronomers do not realize that the dark energy density, $\rho_{DE}$, is not infinitesimal by the standards of the modern laboratory. Recall that the critical energy of the universe is presently taken to be $\rho_c = 9 \times 10^{-10}$ J/m$^3$, and using $\rho_{DE} = 0.70 \times \rho_c$ we obtain

$$\rho_{DE} = 6.3 \times 10^{-10} \text{ J/m}^3$$

and in mass per unit volume $\rho_{DE} \rightarrow 0.70 \times 10^{-26}$ kg/m$^3$ = 3.9 GeV/c$^2$ m$^3$ ~ 4 protons/m$^3$. I use MKS units to emphasize that I am thinking about an experimental question. To think about the magnitude of $\rho_{DE}$, I compare it to an electric field of $E$=1 volt/m. This electric field energy density is

$$\rho_E = \varepsilon_0 E^2/2 = 4.4 \times 10^{-12} \text{ J/m}^3$$

This size electric field is easily detected and measured. Thus in the laboratory and in space we detect and measure electric fields whose energy densities are much less than $\rho_{DE}$. This realization was the impetus for me to consider the possibility of direct detection of $\rho_{DE}$ or its effects.

It is important to compare $\rho_{DE}$ with the energy density of the gravitational field

$$\rho_G = g^2/(8\pi G_N) = 5.7 \times 10^{+10} \text{ J/m}^3 \text{ on the earth's surface}$$

where $g$ is the gravitational acceleration and $G_N$ is Newton's gravitational constant. Thus on the earth's surface $\rho_G$ is enormous compared to $\rho_{DE}$.

### 1.2. Obvious Problems in Direct Detection and Measurement of Dark Energy Density

There are obvious problems in the direct detection and measurement of dark energy density:
- Unlike the electric field the dark energy field cannot be turned off and on.
- We expect that the dark energy field is uniformly distributed; hence we do not expect to find a region with zero dark energy. But it is possible that the distribution is not uniform, that there are regions of larger and smaller density in space.
- In some hypothesis about the nature of dark energy, the dark energy field does not exert a force on any material object. Hence there would be no way to detect it.
- As calculated in the previous subsection, $\rho_G$ at the earth's surface is enormous compared to $\rho_{DE}$, therefore a proposed dark energy detection method that is sensitive to gravity cannot be carried out on the earth's surface. The experiment would have to be carried out in space, sufficiently far from the sun as well as the planets.

## 1.3. Dark Energy and the Planck Scale

Numerous discussions of dark energy are involved with the Planck scale with the so called Planck mass, $M_{Planck}$, and the Planck energy, $E_{Planck}$, given by

$$M_{Planck} = [hc/2\pi G_N]^{1/2} = 1.2\times 10^{19} \text{ GeV}/c^2 = 2.2\times 10^{-8} \text{ kg}$$
$$E_{Planck} = [hc^5/2\pi G_N]^{1/2} = 1.2\times 10^{19} \text{ GeV} = 2.0\times 10^9 \text{ J}$$

Here h is Planck's constant and c is the velocity of light

I am bothered by the prevalent use of $E_{Planck}$ in dark energy discussions. $E_{Planck}$ mixes a quantum mechanical constant with two classical constants, yet ignores another classical constant - the electron charge, $e$, that incorporates the mystery of the quantization of electric charge. Of course it common to say that the Planck energy is the region where quantum mechanics intimately intersects with general relativity and I agree with that vague idea. But what does the Planck energy have to do with dark energy? We do not know if dark energy has anything to do with gravitation. I return to this later when I summarize the infamous problem that arises when dark energy is associated with ground state vacuum fluctuation energy and $E_{Planck}$ is used as a cut-off.

## 1.4. Mass Scales, Length Scales and Dark Energy

### 1.5.1. Review of Mass and Length Scales Concept

An early example of the use of mass and length scales is the astonishing use made by Yukawa to interpret the strong force as caused by pion exchange. We write the strong potential, $V$, as a function of the distance $r$ from the potential source in the simplified form:

$$V = -a \exp(r/L_{strong})$$

where $a$ gives the strength of the potential, and $L_{strong}$ is the range or length scale. The general quantum mechanical relation between $L$ and the mass $m$ of the particle carrying the force is

$$m \times L = h/(2\pi c) \qquad (1)$$

It is convenient to remember that $h/(2\pi c) = 1.97\times 10^{-13}$ MeV/$c^2$ m $= 3.51\times 10^{-43}$ kg m. Thus for the strong force with $L_{strong} = 10^{-15}$ m, we get the pion mass $= 197$ MeV/$c^2$, as Yukawa predicted seventy years ago.

### 1.5.2. Mass and Length Scales for the Weak Interaction

Equation 1 works well for the weak force if we use $m \approx 100$ GeV/$c^2$ to approximate the W and $Z^0$ masses. Then Eq. 1 gives $L_{weak} \approx 2\times 10^{-18}$ m, much smaller than the strong force range as it should be.

### 1.5.3. Mass and Length Scales at the Planck Scale

Using Eq.1 and $M_{Planck} = 2.2\times 10^{-8}$ kg

$$L_{Planck} = 1.6\times 10^{-35} \text{ m}.$$

This length dominates string theory concepts and much theoretical work on quantum gravity

### 1.5.4. Mass and Length Scales for Dark Energy

We do not know a mass $m$ for dark energy to insert in the general quantum mechanical relation of Eq. 1 to calculate $L_{DE}$. We only know the dark energy density, $\rho_{DE}$. Therefore we resort to a dimensional argument,

$$L_{DE} = [hc/(2\pi\rho_{DE})]^{1/4} = 80 \times 10^{-6} \text{ m} = 80\mu\text{m},$$

that may have no validity. But probing dark energy is so difficult that we temporarily accept the idea. $L_{DE}$ corresponds to a frequency

$$f_{DE} = c/L_{DE} = 4 \times 10^{12} \text{ Hz}.$$

Note that in both of these two equations the presence or absence of factors of 2 and $\pi$ are arbitrary.

For later use we write

$$\rho_{DE} = [h/(2\pi c^3)] f_{DE}^4 \qquad (2)$$

## 2. DARK ENERGY AND THE VACUUM ENERGY DENSITY

### 2.1. The Puzzle of the Connection Between Vacuum Energy Density and Dark Energy If There Is a Connection

Conventional quantum mechanics requires that a ground state have non-zero energy. Following Ref. [1], a massless scalar field has ground state energies ½$hf_1$, ½$hf_2$, ½$hf_3$... when the field is treated as an harmonic oscillator. The total vacuum energy density $\rho_{vac}$ is obtained by quantizing the system in a cubic box and integrating up to some cutoff frequency $f_{max}$, yielding

$$\rho_{vac} = [h\pi/(2c^3)] f_{max}^4 \qquad (3)$$

Note that Eqs. 2 and 3 are the same except for factors of 2 and $\pi$.

Eq. 3 leads to an infinite quantity unless $f_{max}$ is finite. Using an argument I don't like because of my mistrust of the significance of the Planck scale, it is conventional to set $f_{max} = f_{Planck} \sim 10^{43}$ Hz. This leads to the result that has been repeated ad nauseam, $\rho_{vac}/\rho_{DE} \sim (f_{max}/f_{DE})^4 \sim 10^{120}$, an unacceptable result that bedevils the idea of a connection between vacuum energy and dark energy.

### 2.2. Some Proposed Cures to the Puzzle

Some proposed cures are:
- Assume that since in reality there are many different fields, their contributions cancel each other exquisitely giving a small net $\rho_{vac}$. An impetus for this concept is the known phenomenon that bosons and fermions have opposite sign contributions to $\rho_{vac}$.
- Find a reason for $f_{max}$ to be much less than $f_{Planck}$.
- Make a dark energy model that has nothing to do with vacuum energy.
- Question the reality of vacuum energy and so disconnect from theories about dark energy [2].

## 3. PROPOSALS FOR DIRECT DETECTION OF DARK ENERGY OR ITS EFFECTS

### 3.1. Superconductor Noise and Dark Energy

Beck and Mackey [3] have proposed (a) that dark energy is only connected to electromagnetic vacuum energy density with a cutoff above $4 \times 10^{12}$ Hz and (b) that this idea can be tested by looking for a decrease above $4 \times 10^{12}$ Hz in the noise spectrum in superconductors. This proposal has been substantially criticized [4].

### 3.2. Dark Energy and Measurements of Gravitational Force at Small Distances

In the past fifteen years there have been several short distance measurements [5-7] of the validity of the inverse square law for the gravitational force. Distances of the order of 50 to 100 μm have been examined. One motivation for these experiments was predictions about the effect of hypothetical string theory extra dimensions on gravity. Another motivation was the possible significance of the dark energy length scale $L_{DE}$ = 80 μm, deduced in Sec. 1.5.4. It is usual to generalize the Newtonian gravitational equation as follows:

$$V(r) = -G_N (m_1 m_2/r) [1+a \exp(-r/\lambda)],$$

where a and λ are parameters for the deviation from Newtonian physics. No deviations have been found [5-7].

### 3.3. Might the Dark Energy Density Be Clumped?

At present my thoughts are concentrated on the possibility that $\rho_{DE}$ is not uniformly distributed, and that the clumping can be detected in space, *sufficiently far away from the gravitational fields of the sun and planets*. The burning question is what can do the detection. For electromagnetic fields we have particles charge, for gravitational fields we have particle mass, but what particle property is sensitive to the dark energy field? This is the fundamental experimental question in dark energy physics. For example, what is the possibility of dark energy producing a phase shift in an atom, and using atom interferometry to detect the phase shifts?

### 3.4. Might There Be A Dark Energy Particle?

In the multitudinous world of dark energy theories the idea that there might be a dark energy particle [8] is an outlier theory. But the experimenter should think even about outliers.

### 3.5. An Experimenter's Conclusion

At the start of the nineteenth century physicists worked on three apparently unrelated phenomena: electricity, magnetism, light. By the end of the nineteenth century these phenomena had been united into electromagnetic theory. At the start of the twenty first century we work on gravitation, dark matter, dark energy, and many problems having to do with the nature of mass. I am optimistic that we will be as productive in fundamental astronomy and physics in the future as we have been in the last two centuries.

### ACKNOWLEDGMENTS

I thank Garrett Morton for his participation in the preparation of this paper. I have had particularly useful conversation on the nature of dark energy with Frans Klinkhamer, Philip Mannheim, Mark Trodden, Thomas Rizzo, Marvin Weinstein and Helen Quinn. This work was supported by Department of Energy contract DE-AC03-76SF00515.